\documentclass[CEJP,PDF]{cej} 
\usepackage{layout}
\usepackage{amsmath}
\usepackage{textcomp}
\usepackage{hyperref}

\begin{document}
\title{In vitro investigation of the effects of X- rays carried by a photon beam upon the cell cycle progression in Vero cells}

\articletype{Research Article}

\author{Cosmin Teodor Mihai\inst{1}$^,$\email{cosmin.mihai@uaic.ro},
        Gabriela Vochita\inst{2},
       Ramona Focea\inst{3},
       Pincu Rotinberg\inst{2}
       }

\institute{
     \inst{1} Department of Biology, "Alexandru Ioan Cuza" University of Iasi,\\
      Bd. Carol I, nr. 20A, 700506,Iasi, Romania
     \inst{2} National Institute of Research and Development for Biological Sciences, branch Institute of Biological Research Iasi,\\
   Str. Lascar Catargi, nr. 47, 700107,Iasi, Romania
     \inst{3} Department of Biophysics, "Alexandru Ioan Cuza" University of Iasi, \\
     Bd. Carol I, nr. 20A, 700506,Iasi, Romania
          }

\abstract{The effects of the X-rays carried out in a photon beam produced by a medical linear particle accelerator on the progression of the cell cycle in normal cells Vero were evaluated by flow cytometric mean. The evaluation of the consequences of 1 and 3 Gy irradiation upon cell cycle development was performed after 96 and 120 hours from the exposure moment, in order to register the late effects of irradiation. As compared with the control group (sham exposed) the treated cells with the dose of 1 Gy have shown an increase of the frequency of cells blocked in S and G2/M stages and also of the apoptotic cells at 96 hours from the treatment. After another 24 hours, the frequency of the blocked cells in S and G2/M has decreased. The response of the 96 hours cell cultures to 3 Gy treatment was faster, as suggested by the increased number of apoptotic cells and by reduced frequency of the arrested cells in S and G2/M. The blockage of Vero cells in different phases of the cell-cycle and the presence of active repairing systems   reveals the reduced risk of the malignant transformation of the exposed normal cells and could offer an enhanced cell protection to another damaging agents.}


\maketitle


\section{Introduction}
Our lifestyle and the consistent presence of the radiation pollution in our environment (result of the large scale use of nuclear energy, of atomic tests for the development of different nuclear weapons, of exposure to cosmic radiations along plane flights, and the natural radiation fund etc.) have elevated the risk of triggering different diseases. Furthermore, the increased number of radiological investigations for a precise and rapid diagnosis, has augmented the exposure to ionizing radiations, which isn't accidental anymore. The long time effects of irradiation are still under investigation \cite{journal-1, journal-2, journal-3}. \\
Radiodiagnosis and radiotherapy are currently used for a rapid and accurate diagnosis and also in the therapy of some illnesses (skin diseases, bone maladies, malignancy etc.). According to the United States National Cancer Institute 50\% of patients with neoplasm are treated with ionizing radiations in some stages of their illness, either alone, or in association with diverse anticancer chemotherapeutics \cite{journal-4}.\\
 In the external beam radiotherapy of the neoplastic disease, the most commonly used device for radiation generation is the linear accelerator (LINAC), which delivers X-rays with different energies to the region of the patient's tumor. Most of X-ray energy goes to tumor, existing concomitantly a minimal scatter of X-ray energy outside the beam. Sharply defined X-ray beam minimizes the side effects of the radiotherapy, so only a small amount of radiations travels to other parts of the body. So, using a LINAC device for the treatment of different types of tumors allows a control of the beam edge sharpness, which improves radiotherapy effectiveness and increases the sparing of the adjacent healthy tissues. Also, LINAC may be programmed to function as a generator of electrons, rather than X – rays, being useful for special oncotherapy Linac \cite{journal-5, journal-6}.\\
 Despite of the technological progress which assures the precise targeting and concentration of the X ray beam in the interested area, the risk of triggering a new carcinogenic spot still persists, even if the quantity of energy transferred to normal cells is small.  It is important to improve the radiotherapeutic protocol (administration schema and doses) in order to optimize the therapeutic efficiency and reduce the side effects \cite{journal-7, journal-8}. Exposure of the cells to the high energy X-rays determines structural and functional lesions of the DNA macromolecules. In this case, the normal cells DNA surveillance and repairing mechanisms are active and either repair these lesions, or directs them on the apoptotic way. These scenarios are the ideally ones, but sometimes the structural damages and functional alterations are not controlled and repaired by the cell monitoring and controlling system and the irradiated cells could be transformed in a cancer generating cell \cite{journal-9}.\\
Among the radiobiological researches, a main direction is represented by the investigation of the X-rays impact upon the genetic material integrity and cell cycle progression, a general feature of this interaction being the blockage of the cells in the G$_2$/M phase \cite{journal-10, journal-11, journal-12}. Cell processes aren't static, the responses to an external stimulus could be immediately (such as transitory blockage of the cells in G$_2$/M phase) and delayed, the last aspect being less investigated, evaluated and documented \cite{journal-13, journal-14, journal-15}. \\
 In the present study, were investigated the effects of X- rays, produced by a medical LINAC device, on cell cycle progression of normal Vero cell line. The work aim was to determine the consequences of different irradiation doses upon the cell cycle progression and to establish the relationship between the moment of exposure and the recovery of cell cycle normal state.\textbf{•}
\section{Materials and methods}
\subsection{Chemicals}
Cell culture reagents, DMEM medium, fetal bovine serum (FBS), trypsin,penicillin and streptomycin were obtained from Biochrom AG (Biochrom AG, Germany). Nuclear Isolation and Stain (NIM-DAPI 10) was obtained from Beckman Coulter (USA).
\subsection{Cell cultures}
The Vero cells (normal kidney monkey cells), ATCC, CCL-81, were cultured in DMEM medium supplemented with $2\%$ fetal calf serum and antibiotics in a $5\%$ $CO_{2}$ atmosphere, at $37^{o}$C. The doubling time of the cell cultures is of approximately 25 hours.
\subsection{High LET X-rays application}
High LET X-rays exposure was carried out in a 6 MV photon beam produced by a linear  particle accelerator VARIAN CLINAC 2100SC type, from the ”St. Spiridon” Emergency County Hospital, Iasi, Romania.
\subsection{Treatment of cells}
The cells were seeded in $25 cm^2$ cell culture flasks at an initial density of 3 x $10^5$ cells /flask. After 24 hours from the initiation of the cell cultures, when the cells have confluenced realizing the monolayer, culture flasks were exposed to the electron flux. \\
The cell cultures were irradiated with 1 Gy and 3 Gy respectively, at a source – sample distance of 98 cm. The dose rate was 260.88 cGy/min. The absorbed doses were calculated at 2 cm depth in the samples, using the following formula \cite{journal-16}.\\

D = $\dot{D}$ $\times$ RDF $\times$ PDD $\times$ t ×0.005029,\\

where D is the prescribed dose in the sample, $\dot{D}$ is the dose rate, at the point of dose maximum on the central axis of the beam, RDF is the relative dose factor and PDD is the percentage depth dose at 2 cm depth, for a 25 x25 cm2 field size, and t is the irradiation time.\\
 The values for $\dot{D}$, RDF and PDD for the given irradiation geometry (25x25 cm$^{2}$ field size, 2 cm depth in the sample) were obtained after beam calibration procedures (in accordance with IAEA TRS-398 dosimetric standard) using a 3D Blue Water Phantom, a PTW Farmer cylindrical ionizing chamber and a PTW Unidose Electrometer.\\
Sham irradiated cells were used in all the experiments as control cells.\\
After irradiation, the cell cultures were allowed to grow for another 48 and 96 hours. \\

\subsection{Cell cycle analysis}
After irradiation, cells were harvested from the surface of culture flasks by trypsinization, resuspended in complete medium and then pelleted by centrifugation for 5 minutes at 1800 rpm. \\
The cells were washed twice in cold PBS. The cell pellet was resuspended in NIM-DAPI (Beckman Coulter, USA) and were stained overnight at 4$^o$C. For every control and treated sample, 20.000 cells were measured on Beckman Coulter Cell LabQuanta SC flowcytometer (Beckman Coulter, USA), using mercury arc lamp 100 W, a 355/37 exciter and a 460 BP filter for the collection of fluorescence and linear amplification.\\
After gating and elimination of the debris, all flow cyotmetric data were collected as LMD files and analyzed using Flowing Software (developed by Cell Imaging Core, Turku Centre for Biotechnology, Abo Akademi University Finland).
\subsection{Statistical analysis}
All of the experiments were carried out with at least three independent repetitions and all data were expressed as the mean value and standard deviation (SD). Statistical analysis was performed using Student's t test and the differences were expressed as significant at level of p$<$0.05.\\
\section{Results}
Cell cycle effects were evaluated by flow cytometry in samples harvested at 96 and 120 hours, respectively, from the treatment with the doses of 1 and 3 Gy of X-rays carried by a photon beam.

\begin{figure}[h]%
\centering
\parbox{2.5in}{\includegraphics[width=2.5in]{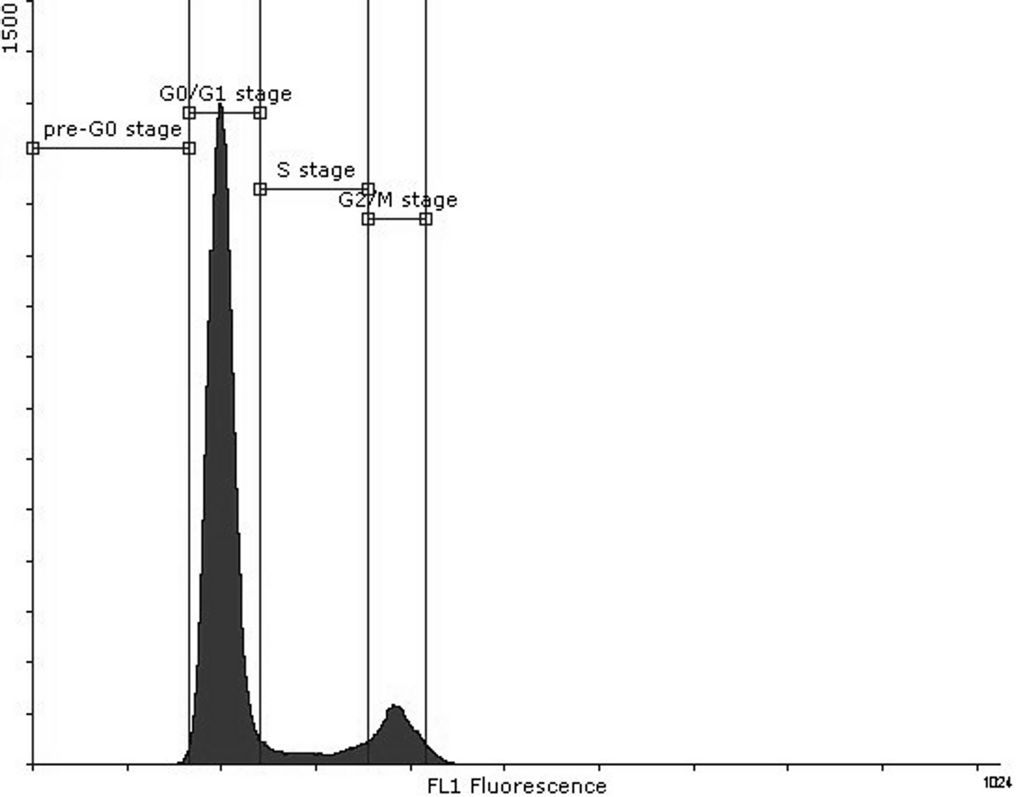} }%
\qquad
\begin{minipage}{2.5in}%
\includegraphics[width=2.5in]{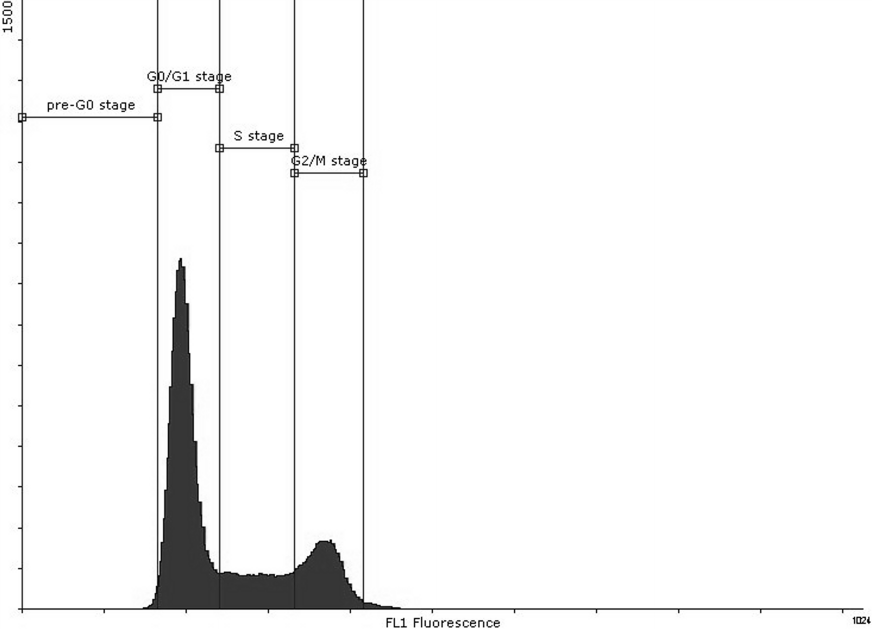}
\end{minipage}%
\caption{Gated population of Vero cells after doublets and debris removal, and the specific markers for the cell cycle stages. The profiles showed dual-variable plots of cell number versus DAPI uptake. pre-G$_0$, G$_0$/G$_1$, S, and G$_2$/M cell populations were quantified. In the histograms are depicted the corresponding gates to pre-G$_0$, G$_0$/G$_1$, S and G$_2$/M phases.Left image - control group corresponding to 96 hours, right image - control group corresponding to 120 hours.\label{fig1}}
\label{fig:1figs}%
\end{figure}

\begin{table}[!h]
\caption{Impact of high energy transfer X- rays in doses of 1 Gy and 3 Gy upon the percentage of single Vero cells in different stages of the cell cycles. The pre-G0 stage cells were also included as an indicator for the presence of the apoptotic cells.\label{tab1}}
\begin{tabular}{ccccc}
\hline
      & Pre-G$_0$ stage & G$_0$/$G_1$ stage & S stage & G$_2$/M stage\\
		Samples & \%Mean$\pm$SEM & \%Mean$\pm$SEM & \%Mean$\pm$SEM & \%Mean$\pm$SEM \\ 
     \hline \hline
 \multicolumn{5}{c}{96 hours} \\ \cline{1-5}
Control  &  0.0983$\pm$0.0015  &  84.50$\pm$1.27 &  4.35$\pm$0.07 &  11.50$\pm$0.17 \\
1 Gy treated &  0.1739$\pm$0.0026 & 80.95$\pm$1.22 & 6.31$\pm$0.10 & 12.68$\pm$0.19 \\
3 Gy treated &  0.4814$\pm$0.0073 & 83.77$\pm$1.26 &  4.67$\pm$0.07 & 11.94$\pm$0.18 \\
\hline \hline
\multicolumn{5}{c}{120 hours} \\
\cline{1-5}
Control&0.26$\pm$0.01&63.65$\pm$0.96&16.43$\pm$0.25&20.17$\pm$0.30 \\
1 Gy treated&0.55$\pm$0.02&73.15$\pm$1.10&11.90$\pm$0.18&15.25$\pm$0.45\\
3 Gy treated&0.52$\pm$0.01&75.31$\pm$1.14&12.18$\pm$0.18&12.81$\pm$0.19 \\
\hline

\end{tabular}
\paragraph{}
Errors indicate the standard error of the mean (SEM) for n=3 independent experiments.
\end{table}

\begin{figure}[h]%
\centering
\parbox{3in}{\includegraphics[width=3in]{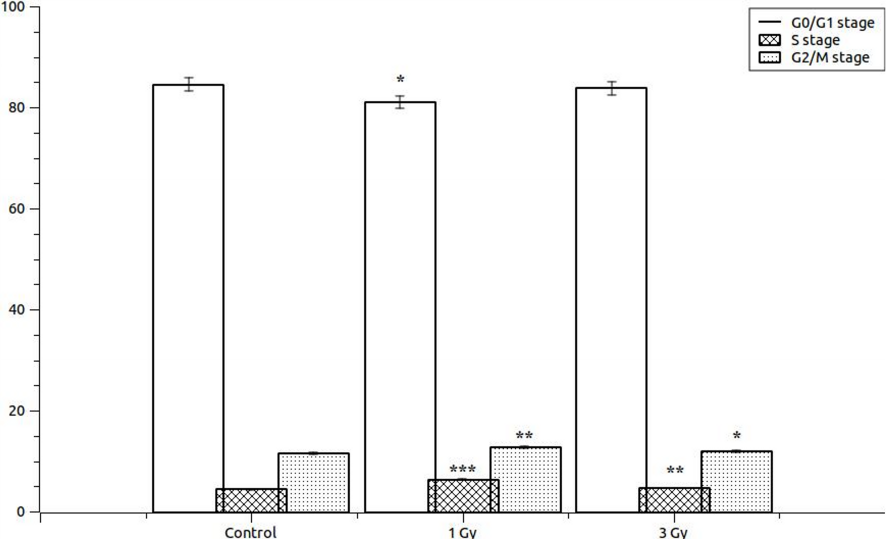} }%
\qquad
\begin{minipage}{3in}%
\includegraphics[width=3in]{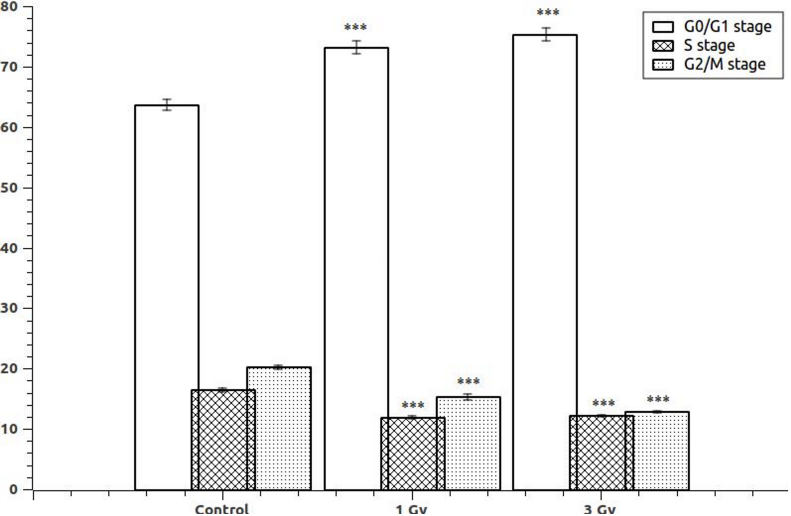}
\end{minipage}%
\caption{Percent distribution of the cells in G$_0$/G$_1$, S and G$_2$/M phases after 96 hours (left image) and 120 hours (right image) after the treatment with low energy transfer x rays (1 Gy and 3 Gy). Error bars indicate the standard error of the mean (SEM) for n=3 independent experiment.\label{fig2}}
\label{fig:1figs}%
\end{figure}

At 96 hours from the treatment, 1 Gy treated cells presented an accumulation (45.17\%) of the cells in phases S and G$_2$/M (with 10.3\%). After 120 hours from the treatment, the fraction of the cells in S and G$_2$/M phases decreased significantly (with 27.55\% and 25.35\% respectively).\\
3 Gy treatment has determined, after 96 hours, minor increases of cells frequency in S (with 7.28\%) and G$_2$/M (3.85\%) stages. After 120 hours, the frequency of the cells in S (with 25.85\%) and G$_2$/M (36.49\%) stages decreased significantly. \\
The frequency of the apoptotic cells, shown in the same table as pre-G$_0$0 phase, wasn't high. 
Compared to control group the treatment has induced after 96 hours an increase of the apoptotic cells number with 76\% in the case of the 1 Gy exposure and with 389\%, in the case of the 3 Gy dose. Thus, is revealed the existence of an active apoptotic processes and its dependence on irradiation dose. After 120 hours exposure, the level of apoptotic cells percentage was maintained at a high level (211\% and 200\% respectively), but the differences between 1 and 3 Gy respectively treatments weren't so clear.\\

\section{Discussions}
The extreme precision and reliability over countless generations of duplication and division of the cells are sine qua non conditions for the development of an organism. To assure the correctness of the cell cycle events eukariotic cell contains a complex regulatory network – called the cell-cycle control system – that controls their timing and coordination. This control system is essentially a robust and reliable biochemical timer that is activated at the beginning of a new cell cycle and is programmed to switch on cell cycle events at the correct time and in the correct order \cite{journal-17}. One of the main features of this system is its independence from the events that is controlling. However, the order and alternation of cell cycle events are reinforced by the dependence of one event on another and by feedback from cell-cycle machinery to the control system. If the control system detects problems in the completion of an event, it will delay the initiation of later events until those problems are solved. Sometimes, the defects in cell cycle controlling machinery could be responsible for the initiation and development of transformations causing the malignant disease \cite{journal-9, journal-17, journal-18, journal-19}.\\
The effects of high energy transfer X-rays are transitory, the induced structural modifications being identified and repaired. The late progression of the specific events, which are responsible for the stability of the cell, as a dynamic system, aren't enough prospected.\\
In order to evaluate the late effects of the exposure to X rays produced by a LINAC, upon Vero normal cells it was investigated their progression through cell cycle and their distribution in different stages of cell cycle. 
Thus, the 96 hours cells treated with 1 Gy high linear energy transfer X-rays were blocked in the S and G$_2$/M stages of the cell cycle. The blockage of the cells in S and G$_2$/M phases is due to errors in the structure of DNA generated by X rays. Also, the progression of the cells through cell cycle is accompanied by an increase of the apoptotic cells numbers (revealed by the left peak from G$_0$/G$_1$), as compared with control cells. This fact suggests that the functionality of reparatory mechanisms aren't affected by the irradiation, determining the stop of cells in S and G$_2$/M phases and driving them to the death by apoptosis. After another 24 hours, the restrictions imposed by the controlling mechanisms, have reduced both the frequency of the cells in S and G$_2$/M phases and the dividing pool of cells, suggesting either their arrest in the G$_0$ state or the augmentation of apoptosis process of the cells with irreparable DNA alterations\cite{journal-20}. \\
The response of the 96 hours cell cultures to 3 Gy treatment is faster, suggested by the increased number of apoptotic cells and by reduced frequency of the arrested cells in S and G$_2$/M stage. The treatment with a dose of 3 Gy has presented, after 120 hours, similar characteristics with the 1 Gy irradiated cell cultures, the proportion of the cells specific to every stage being somewhat modified.\\
The recovery of the cells to the normal state is dependent by the amount of radiation energy transferred to the cell. Thus, the 1 Gy dose generates reparable DNA structural changes, because the cell reparatory mechanism of the nuclear material integrity maintains the fidelity of the DNA biosynthesis by arresting the cell cycle before G$_2$/M. Comparatively with 1Gy radiations, the 3 Gy effects are more pronounced and occurs earlier.\\ 
 According to recent investigations \cite{journal-21, journal-22}, the action of the low doses of ionizing radiations could protect the cells to the mutagenic action of others agents (hormesis phenomenon), especially of the high energy radiations. In our experimental conditions, the blockage of Vero cells in different phases of the cell-cycle and the presence of active repairing systems further argues the enhanced cell protection to other damaging agents.\\
 
\acknowledgements
This study was possible with financial support from the Sectoral Operational Programme for Human Resources Development, project “Developing the innovation capacity and improving the impact of research through post-doctoral programmes”, POSDRU/89/1.5/S/49944 and POSDRU/88/1.5/S/47646

\end{document}